\documentclass[aps,prbbib,twocolumn,epsf]{revtex4}

\usepackage{graphicx}
\usepackage[usenames]{color}
\begin{document}
\draft

\title{Electron beam induced current in the high injection regime}
\author{Paul M. Haney$^1$, Heayoung P. Yoon$^{1,2}$, Prakash Koirala$^3$, Robert W. Collins$^3$,  Nikolai B. Zhitenev$^1$}
\affiliation{1.  Center for Nanoscale Science and Technology, National Institute of Standards and Technology, Gaithersburg, MD 20899 \\
2.  Maryland NanoCenter, University of Maryland, College Park, MD 20742, USA \\
3.  Department of Physics and Astronomy, University of Toledo, Toledo, OH, 43606, USA}

\affiliation{Center for Nanoscale Science and Technology,
National Institute of Standards and Technology, Gaithersburg,
Maryland 20899-6202, USA }
\begin{abstract}
Electron beam induced current (EBIC) is a powerful technique which measures the charge collection efficiency of photovoltaics with sub-micron spatial resolution.  The exciting electron beam results in a high generation rate density of electron-hole pairs, which may drive the system into nonlinear regimes.  An analytic model is presented which describes the EBIC response when the {\it total} electron-hole pair generation rate exceeds the rate at which carriers are extracted by the photovoltaic cell, and charge accumulation and screening occur.  The model provides a simple estimate of the onset of the high injection regime in terms of the material resistivity and thickness, and provides a straightforward way to predict the EBIC lineshape in the high injection regime.  The model is verified by comparing its predictions to numerical simulations in 1 and 2 dimensions.  Features of the experimental data, such as the magnitude and position of maximum collection efficiency versus electron beam current, are consistent with the 3 dimensional model.
\end{abstract}

\maketitle

\section{Introduction}

Thin-film polycrystalline photovoltaics are a fully mature technology, with high power conversion efficiency (21 \% for CdTe), low fabrication costs, and proven competitiveness in the photovoltaic market \cite{kumar}.  However, despite decades of research, basic questions persist regarding the role of microstructure in their operation and performance \cite{bosio}.  One reason for this is that many established techniques for photovoltaic characterization, such as luminescense or charge transport methods, probe the material properties on length scales far greater than that of the material structural and electronic inhomogeneity (e.g. the grain size - typically 1 ${\rm \mu m}$).  Electron beam induced current (EBIC), on the other hand, is a measurement technique which can spatially resolve the electrical response of the material on length scales less than a grain size \cite{Hanoka}.  In an EBIC experiment, a focused beam of high energy electrons impinges on a sample, generating free electron-hole pairs.  The size of the electron-hole pair generation bulb depends on the electron beam energy: e.g. for a beam energy of 3 keV, the excitation bulb length scale is 50 nm in CdTe \cite{gruen}.  Some fraction of the excited electron-hole pairs are separated by internal fields and extracted, resulting in a measured charge current.  The ratio of collected current to charge generation rate is the EBIC efficiency, and can be used to characterize the recombination.  EBIC is naturally suited to measure the minority carrier diffusion length and the surface recombination velocity \cite{wu,roosbroek,donolato}.  Recent work also studies grain boundary properties using EBIC \cite{yoon}, identifying ${\rm CdCl_2}$ treatment as the key to changing grain boundaries from regions of increased recombination to regions of increased charge collection \cite{li,modes}.

Despite the advantages of EBIC measurements described above, quantitative analysis in thin film polycrystalline materials is challenging: basic properties of the data (such as the maximum collection efficiency) do not conform with established models of EBIC response.  Ref. \cite{Nichterwitz} presents a critical analysis of EBIC models, and uses numerical simulation to demonstrate two scenarios which lead to deviations from the ``expected" behavior of EBIC signals.  In the first scenario, a diffusion length less than the drift length results in a reduced collection efficiency within the depletion region.  Our recent work develops an analytical model for describing EBIC response in this regime \cite{haney}.  In the second scenario, large electron beam currents result in ``high injection" of electron-hole pairs, drastically changing the electrostatic potential and the resulting EBIC lineshape.  In this work, we explore the effect of high injection with a combination of numerical and analytical models in 1, 2, and 3 dimensions.  High injection effects have generally been considered important when the density of injected charge exceeds the doping density $N$ \cite{sze}.  We find that the onset of the high injection regime is more specifically related to the product $N \mu V_{\rm bi}\left(L\right)^{dim-2}$, where $\mu$ is the majority carrier mobility, $V_{\rm bi}$ is the built-in potential, $L$ is the sample thickness, and $dim$ is the system dimensionality.  Roughly speaking, when carriers are generated at a rate which exceeds the maximum current accommodated by ``built-in" electric fields and material resistivity, charges accumulate and screen the built-in field.  This results in major distortions in the built-in potential and subsequent changes of the EBIC signal.

The high injection regime may be accessible in an EBIC experiment due to the large number of electron-hole pairs generated per incident electron.  Each electron generates approximately $E_{\rm beam}/\left(3\times E_g\right)$ electron-hole pairs, where $E_g$ is the material bandgap and $E_{\rm beam}$ is the energy of the exciting electron beam (typically $>{\rm 3~keV}$).  For a material with $E_g=1.5~{\rm eV}$, and a beam current of $200~{\rm pA}$ with energy $5~{\rm keV}$, the excitation bulb volume is $V_{\rm b}\approx \left(100~{\rm nm}\right)^3$.  The resulting excitation rate density is $10^{26}~{\rm cm^{-3}~s^{-1}}$, exceeding more typical generation rate densities of $10^{21}~{\rm cm^{-3}~s^{-1}}$ (1 sun illumination) by a factor of $10^5$.  This indicates that EBIC experiments may drive the system into a nonlinear regime.  Experiments typically strive to ensure the system remains in the linear regime, in part because a theoretical treatment of the material response in nonlinear regimes is less well developed.  The current work aims to describe one aspect of nonlinear response, specifically charge accumulation and resulting screening of internal fields in the material.  We find an implicit expression for the EBIC response as a function of total generation rate and beam position.  We then compare the model predictions to experimental results of CdTe, and find qualitatively similar features.

The paper is organized as follows: in Sec. II we describe the numerical model in general terms, and describe the analytical model in 1, 2 and 3 dimensions.  We compare the analytical model to the numerical simulation in 1 and 2 dimensions.  We then present comparisons between EBIC experiments on two samples and the 3 dimensional model.  We formulate the onset of high injection effects in terms of the device thickness, the built-in potential, and the absorber resistivity (recall the resistivity $\rho$ is given by $\rho=\left(q\mu N\right)^{-1}$, where $q$ is the absolute value of the electron charge, $\mu$ is the carrier mobility, and $N$ is the carrier density (doping)).

\section{Model} \label{sec:model}

To develop a physical picture of the system response in the high injection regime, we use numerical simulation to assist in identifying the key physics involved and to serve as a check on the analytical models.  The numerical model is a standard implementation of the coupled drift-diffusion equations for electrons and holes combined with the Poisson equation, and includes Shockley-Read-Hall recombination.  We choose parameters that are typical for thin film solar cells (see the caption of Fig. \ref{fig:1d}).

We also preface the discussion with a comment on the role of system dimensionality.  In the classic description of EBIC measurements by Donolato and others, the collection efficiency is governed by the diffusion and recombination of minority carriers in the neutral region.  These works explicitly show that, for the purposes of calculating the EBIC efficiency, the 3-dimensional system can be reduced to a 1-d system \cite{donolato}.  This simplification follows from the system linearity, and the fact that the quantity of interest (the EBIC efficiency) is a {\it ratio} of the system response (the current) to the driving input (the generation rate).  We consider nonlinear effects in this work and find that the system behavior changes qualitatively with system dimensionality.  Experimental comparisons are only meaningful for the 3-dimensional model, nevertheless the systems of reduced dimension offer clear insight into the important physics which govern this nonlinear regime.

We present results in terms of the EBIC efficiency $\eta$, defined as the ratio of the measured current to the total charge generation rate of electron-hole pairs $G$.   The experimental generation rate depends on beam energy, and is estimated as \cite{wu}:
\begin{eqnarray}
G = \left(1-b\right)\frac{\left(I_{\rm beam}/q\right)\times \left(E_{\rm beam}/E_0\right)}{3 \times \left(E_g/E_0\right)}, \label{eq:G}
\end{eqnarray}
Note that the units of $G$ depend on the system dimensionality $dim$ , according to $G\propto{\rm s^{-1} m}^{dim-3}$.  Also note that $G$ is is the {\it total} generation rate (as opposed to the generation rate {\it density}).  The role of dimensionality in the system behavior is discussed at length in Sec. \ref{sec:3dmain}.  Additionally, the models presented here assume an excitation length scale which is smaller than the depletion width and diffusion length, so that they apply most directly to low beam energies (lower than $5~{\rm keV}$ for materials such as CdTe).


\subsubsection{1-d analysis}

\begin{figure}[h!]
\begin{center}
\vskip 0.2 cm
\includegraphics[width=3.5in]{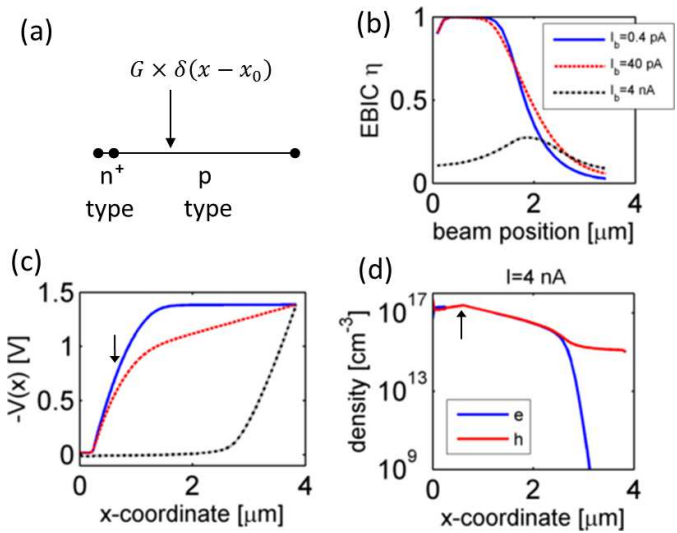}
\vskip 0.2 cm \caption{(a) 1-d system geometry.  The excitation is taken to be a delta function.  (b) The EBIC lineshape for three values of the electron beam current.  (c)  The band diagrams corresponding to the three electron beam currents from (b), for an excitation position shown by the arrow.  (d)  The nonequilibrium electron and hole density (log scale) for the highest beam current, for an excitation position indicated by the arrow.  The electron and hole density are approximately equal over the region of the screening of the built-in field.  We take $\mu=1~{\rm cm^2/V\cdot s},~\tau=10~{\rm ns}$ (identical values for electrons and holes), $N_A=10^{15}~{\rm cm^{-3}},~N_D=5\times 10^{16}~{\rm cm^{-3}}$, $p$ and $n$-type region thicknesses are $3.6~{\rm \mu m}$ and $0.25~{\rm \mu m}$, respectively.  ${\epsilon}=12~\epsilon_0,~E_g=1.5~{\rm eV}$, where $\epsilon_0$ is the permittivity of free space.}\label{fig:1d}
\end{center}
\end{figure}

We begin with a 1-d model of a $n^{+}$-$p$ junction, depicted in Fig. \ref{fig:1d}(a).  Fig. \ref{fig:1d}(b) shows the EBIC lineshapes for increasing values of the electron beam current \cite{footnote1}.  For lower values of the beam current, the lineshape consists of a plateau of nearly perfect collection efficiency over the width of the depletion region, followed by an exponential decrease of the collection into the neutral region.  The length scale of the EBIC signal decay is the minority carrier diffusion length.  This regime is well described by the previously derived analytic models \cite{donolato}. For larger beam currents, the maximum collection efficiency is decreased, and is shifted into the interior of the sample.  Fig. \ref{fig:1d}(c) shows the electrostatic potential in the device for three values of the beam current, for a delta-function excitation located deep within the depletion region at $x=0.6~{\rm {\mu m}}$.  For low beam current, the energy bands are only slightly perturbed from their equilibrium values.  At higher beam currents, an increasing portion of the potential drop across the device takes place over the neutral region.  This potential drop can be understood as the driving force for {\it majority} carriers from the $p$-type side of the device: the potential drop over the neutral region $\Delta V$ satisfies $G=\mu N_A \Delta V /L_{\rm neutral}$.  Here $\mu$ is the hole mobility, $N_A$ is the doping density, and $L_{\rm neutral}$ is the length of the neutral region.  As $G$ increases, the potential drop required to extract all the carriers exceeds the built-in potential.  At this point, charge accumulation sets in, which partially screens the built-in field.  This is shown in the curve of Fig. \ref{fig:1d}(c) for $I=4~{\rm nA}$ (black dashed curve).  Carriers diffuse and recombine within this screened region, and the potential drop occurs over a smaller length, increasing the driving field.  In this regime, the collection efficiency is reduced.

The physical picture described here is readily formulated with a model.  We first describe the model for an excitation deep within the depletion region at a position $x_B$, and assume that the spatial extent of the excitation is smaller than other length scales of the problem (e.g. diffusion length and depletion width), and can therefore be described by a delta function positioned at $x_B$.  If the total generation rate drives the system into the high injection regime, the built-in field is screened over a length scale $L^*$.  Within this screened region, charges diffuse and recombine.  The continuity equation for holes is:
\begin{eqnarray}
D \frac{\partial^2 p}{\partial x^2}=\frac{p}{2\tau}. \label{eq:dd1d}
\end{eqnarray}
where $D$ is the hole diffusivity, and $\tau$ is the carrier lifetime (we assume both quantities are equal for electrons and holes).  The factor of 2 in the denominator of the right hand side of Eq. (\ref{eq:dd1d}) follows from the form of the Shockley-Read-Hall recombination: when the density of electrons and holes are equal (which applies here, as shown in Fig. \ref{fig:1d}(d)), the effective lifetime is $2\tau$ \cite{footnote2}.

The first boundary condition on $p(x)$ is a discontinuity in the current $j$ at the excitation position $x_B$.  For the second boundary condition, we note that the field sweeps carriers out at the edge of the screened region (at a position $L^*$), so that $p(L^*)=0$.  For the final boundary condition, we let the carrier current vanish at $x=0$, as hole carriers are not collected by the $n$-type contact.  The mathematical formulation of the boundary conditions and the resulting solution for $p(x)$ is given in Appendix A.

Given $p(x)$, we can compute the total recombination which occurs in the screened region:
\begin{eqnarray}
R_{\rm tot} &=& \int_{0}^{L^*}  \frac{p(x)}{2\tau} dx\\
&=& G\left(1-\cosh\left(\frac{x_B}{L_D'}\right){\rm sech}\left(\frac{x_B+L^*}{L_D'}\right)\right) \label{eq:R1d}
\end{eqnarray}
Here $L_D'=\sqrt{2D\tau}$ is the effective diffusion length.  The potential drop $V_{\rm bi}$ is confined to the region outside of $L^*$, and therefore takes places over a length $L-L^*$.  The majority carrier current (i.e the hole current) is therefore given by:
\begin{eqnarray}
J&=&\frac{\mu N_A V_{\rm bi}}{L-L^*}. \label{eq:j1d}
\end{eqnarray}

The total generation rate is equal to the sum of the total recombination and the collected current: $G=R_{\rm tot}+J$.  Written in terms of Eqs. (\ref{eq:R1d}-\ref{eq:j1d}), this identity takes the form:
\begin{eqnarray}
\frac{\mu N_A V_{\rm bi}}{L-L^*}&=&G \cosh\left(\frac{x_B}{L_D'}\right){\rm sech}\left(\frac{x_B+L^*}{L_D'}\right)\label{eq:1d}
\end{eqnarray}
The above equation determines the size of the screened region $L^*$ for a given value of $G$.  Once $L^*$ is determined, the current can be evaluated directly from Eq. (\ref{eq:j1d}), giving the EBIC efficiency.

Letting $L^*\rightarrow 0$ corresponds to the onset of the high injection regime.  Eq. (\ref{eq:j1d}) provides an estimate of the maximum current which can be accomdated by the material before charge accumulation and screening sets in.  This in turn determines a critical generation rate $G_{\rm crit}$, above which high-injection effects reduce the maximum EBIC efficiency:
\begin{eqnarray}
G_{\rm crit}^{{\rm 1d}}=\frac{\mu N_A V_{\rm bi}}{L} \label{eq:1dcrit}
\end{eqnarray}

\begin{figure}[h!]
\begin{center}
\vskip 0.2 cm
\includegraphics[width=3.4in]{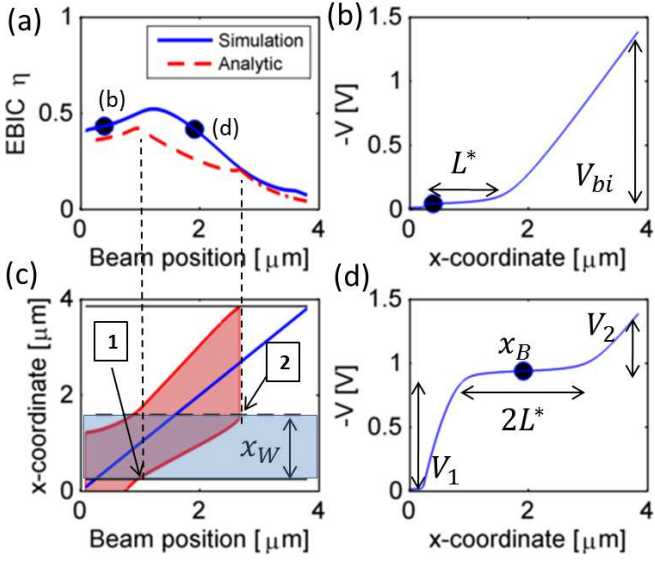}
\vskip 0.2 cm \caption{(a) shows the EBIC lineshape under high injection conditions.  (b) and (d) show the electrostatic potential profiles at the different beam positions indicated on the curve in (a).  The location of the screened region is shifted according to the beam position $x_B$.  The model is modified accordingly, as described in the text.  (c) shows the evolution of the screened region within the device as the beam position is varied.  The y-axis represents the real-space coordinate within the device.  The x-axis is denotes the beam position.  The diagonal blue line shows the position of the beam within the device, while the red shaded region area the screened region within the device.  The depletion region is also indicated in blue.  The specially labelled points ``1" and ``2" denote where kinks occur in the analytic EBIC plot in (a).}\label{fig:1dtrace}
\end{center}
\end{figure}

So far we've assumed that the excitation is located within the depletion region, so that the charge accumulation and screening is localized within the built-in field, as in Fig. \ref{fig:1dtrace}(b).  We next describe the EBIC response for a general beam position $x_B$.  As the beam position is varied, the region of charge accumulation and screening moves away from the depletion region.  This is shown schematically in Fig. \ref{fig:1dtrace} (c), which shows how the area of the screened region in the device (red shaded region along the y-axis) varies with the beam current position (which varies along the x-axis).  The changing position of the screened region has two effects: the first is that the screened region may no longer occupy the entire depletion width, so that some of the equilibrium potential drop re-appears there (shown as $V_1$ in Fig. \ref{fig:1dtrace}(d) - notice that $V_1$ is the equilibrium potential $V_{\rm eq}$ evaluated at $x=x_B-L^*$).  This in turn decreases the potential drop driving the majority carriers out of the neutral $p$ region ($V_2$ in Fig. \ref{fig:1dtrace}(d) - this potential drop $V_2$ is given by $V_2=V_{\rm bi}-V_1$).  This effect is incorporated by modifying the potential appearing in Eq. \ref{eq:j1d}:
\begin{eqnarray}
V_{\rm bi} \rightarrow V_{\rm bi}-V_{\rm eq}\left(x_B-L^*\right) \label{eq:vbi_new}
\end{eqnarray}
where the equilibrium potential $V_{\rm eq}\left(x\right)$ in the depletion region is of the familiar form:
\begin{eqnarray}
V_{\rm eq} (x)=\frac{N_a}{2\epsilon}\left(x-x_W\right)^2
\end{eqnarray}
Here $x_W$ is the depletion width, given by $x_W=\sqrt{2 V_{\rm bi}/(q \epsilon N_A)}$.  Note that Eq. \ref{eq:vbi_new} is only applicable for $x_B-L^*>0$.

The second effect of changing the beam position is that the length over which $V_2$ drops is decreased.  This is incorporated by decreasing the length of the the potential drop appearing in Eq. \ref{eq:j1d}, or equivalently decreasing the device thickness $L$ according to
\begin{eqnarray}
L \rightarrow L-x_B
\end{eqnarray}

Incorporating both modifications in Eq. \ref{eq:1d} gives a nonlinear equation for the screening length $L^*$ for a given beam position $x_B$:
\begin{eqnarray}
\frac{\mu N_A \left(V_{\rm bi}-\frac{N_A}{2\epsilon}\left(x_B-L^*-x_W\right)^2\right)}{L-L^*-x_B}&=& \nonumber\\G \cosh\left(\frac{L^*}{2L_D'}\right){\rm sech}\left(\frac{L^*}{L_D'}\right)\label{eq:1dtrace}
\end{eqnarray}

As before, for a given $G$ and $x_B$, Eq. \ref{eq:1dtrace} can be solved for $L^*$, from which the EBIC efficiency is determined by plugging this value into Eq. \ref{eq:j1d}.  For beam positions sufficiently far away from the depletion region, the screened region is fully separated from the depletion region (see point "2" in Fig. \ref{fig:1dtrace}(c)).  At this point, we let the EBIC efficiency decay with increasing distance according to the effective diffusion length.

\begin{figure}[h!]
\begin{center}
\vskip 0.2 cm
\includegraphics[width=3.5in]{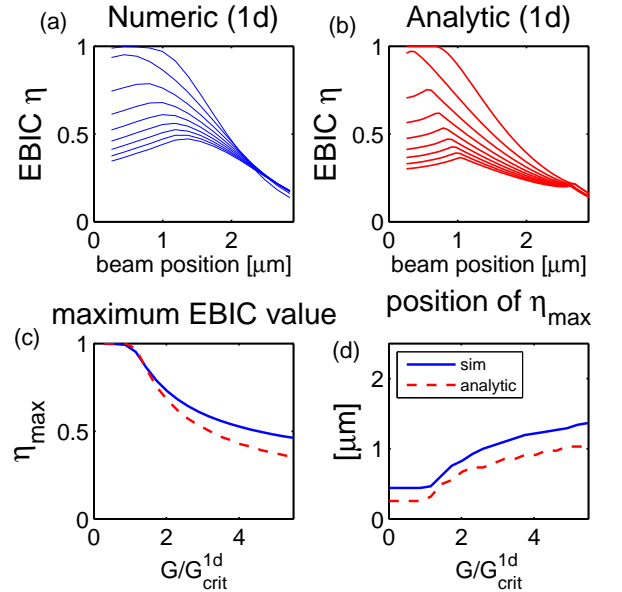}
\vskip 0.2 cm \caption{(a) shows the simulated EBIC lineshapes for a series of total generation rates, which vary from $0.008\times G_{\rm crit}^{1d}$ to $8.6\times G_{\rm crit}^{1d}$ in 20 equal steps.  (b) shows the corresponding analytical model predictions.  (c) shows the maximum EBIC value for simulation and analytical models versus total generation rate.  (d) shows the position of the EBIC maximum for simulation and analytical models as a function of total generation rate.}\label{fig:ebic1d}
\end{center}
\end{figure}

Eqs. \ref{eq:j1d} and \ref{eq:1dtrace} lead to an estimate of the EBIC response versus beam position in the high injection regime.  A comparison of this model with full numerical results are shown in Figs. \ref{fig:ebic1d}(a) and (b).  Overall the lineshapes show qualitative similarities.  Generally the analytical model exhibits two characteristic ``kinks".  At higher generation rates, the first kink is located at the maximum EBIC value.  This first kink corresponds to the point at which the screened region no longer covers the entire depletion width (the point labeled ``1" in Fig. \ref{fig:1dtrace}(c)).  At this point, the majority carrier driving potential is the full $V_{\rm bi}$, while the length over which this potential drops is minimized.  For beam positions greater than this, the driving potential decreases, and the collection efficiency also decreases.  The second kink corresponds to the point at which the screening region no longer intersects any portion of the depletion width (the point labeled ``2" in Fig. \ref{fig:1dtrace}(c)).

Fig. \ref{fig:ebic1d}(c) shows the value of the maximum EBIC efficiency versus total generation rate (scaled by the critical generation rate $G_{\rm crit}^{1d}$, see Eq. \ref{eq:1dcrit}) for the analytical model and numerical simulation.  We find good qualitative agreement, although the analytical model underestimates the EBIC efficiency slightly.  Fig. \ref{fig:ebic1d}(d) shows the position of the maximum EBIC efficiency versus total generation rate.  Here the analytical model captures the trend, but predicts a maximum position which is systematically smaller than found in the numerical simulation.  The shift of the maximum EBIC position towards the center of the device with increasing generation rate was also found in Ref. \cite{Nichterwitz}, and is a distinguishing feature of EBIC in the high injection regime.

\subsubsection{2-d analysis}

\begin{figure}[h!]
\begin{center}
\vskip 0.2 cm
\includegraphics[width=3.5in]{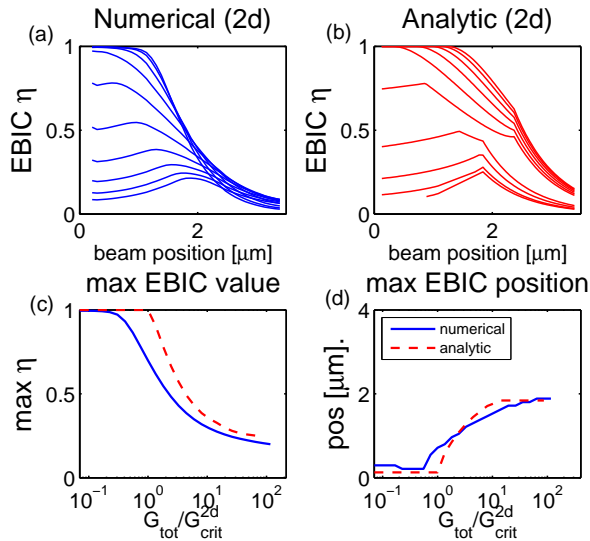}
\vskip 0.2 cm \caption{(a) simulation results for EBIC lineshape in 2-dimensional model, for a series of 9 generation rates from $0.03\times G_{\rm crit}^{2{\rm d}}$ to $39\times G_{\rm crit}^{2{\rm d} }$ (equally spaced on a log scale).  (b) shows the analytical model prediction for the lineshapes for the same parameters.  (c) shows the maximum EBIC signal versus generation rate (solid line is simulation, dotted line is analytical model), and (d) shows the position of the EBIC maximum versus generation rate for simulation and analytical models.  }\label{fig:ebic2d}
\end{center}
\end{figure}

\begin{figure}[h!]
\begin{center}
\vskip 0.2 cm
\includegraphics[width=2.5in]{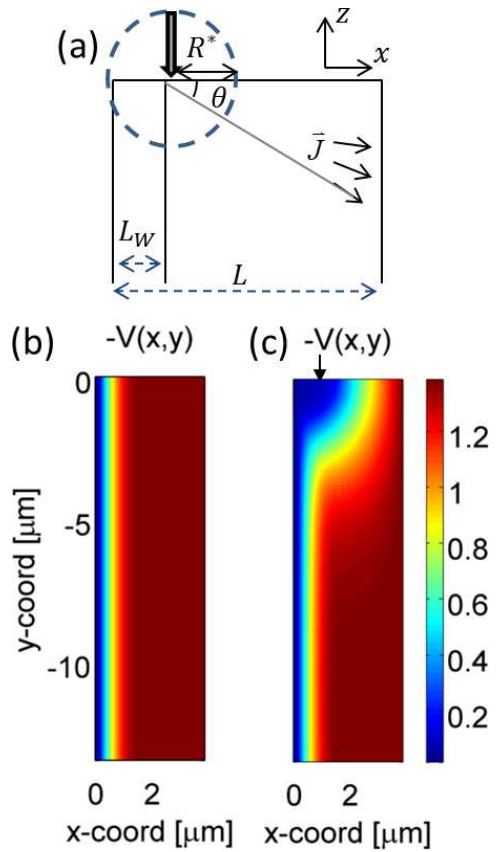}
\vskip 0.2 cm \caption{(a) the cartoon for the analytic model in 2-d.  (b) the equilibrium electrostatic potential, (c) the electrostatic potential under high injection conditions.  This informs the construction of the model depicted in (a).  The arrow indicates the excitation position.  We use the same parameters as given in the caption of Fig. \ref{fig:1d}. }\label{fig:2d}
\end{center}
\end{figure}

We next consider a system in 2-dimensions.  We perform simulations for a two-dimensional system with the same parameters as the 1-d case.  Fig. \ref{fig:ebic2d}(a) shows that, as before, the maximum EBIC signal is reduced as the beam current and resulting generation rate increase past a certain threshold.  Fig. \ref{fig:2d}(b) and (c) show the electrostatic potential in equilibrium and under high injection conditions, respectively.  Under high injection conditions, the potential is screened as in the 1-d case, except the screening is localized near the excitation at the surface.

Extending the physical picture described in the previous section to systems with higher dimension is straightforward.  As before, we begin by assuming an excitation positioned deep within the depletion width.  The charge accumulation screens the internal field over a length scale of $R^*$ (see Fig. \ref{fig:2d}(a)).  Within this screened region, charges diffuse and recombine.  The built-in potential drop occurs outside the screened region and drives out majority carriers.  We compute the total recombination and total extracted current, and equate their sum to the total generation rate.  The resulting identity takes the form:
\begin{widetext}
\begin{eqnarray}
G&=&\frac{-i\pi G R^*}{2L_D'} \left\{\frac{Y_0\left(-iR^*/L_D'\right)}{J_0\left(iR^*/L_D'\right)}J_1\left(-iR^*/L_D'\right) + \left(-iR^*/L_D'\right) + \frac{2L_d}{\pi R^*}\right\}-\frac{2\mu N V_{\rm bi}}{\sqrt{R^*/L-1}}\tanh^{-1}\left(\frac{1+R^*/L}{\sqrt{R^*/L-1}}\right), \label{eq:2debic}
\end{eqnarray}
\end{widetext}
where $J_0$ ($Y_0$) is the Bessel functions of the first (second) kind, and $L$ is the device thickness.  In the above, the left-hand side is the total generation rate, the term in brackets on the right-hand side is the total recombination, and the last term on the right-hand side is the total extracted current.  The detailed mathematical derivation of this can be found in Appendix B.  Given a generation rate $G$, Eq. \ref{eq:2debic} is solved for the screening length $R^*$, from which the EBIC efficiency is readily computed.  For a general beam position $x_B$, we make the same replacements in Eq. \ref{eq:2debic} for $V_{\rm bi}$ and $L$ as in the 1-d case:
\begin{eqnarray}
V_{\rm bi} &\rightarrow& V_{\rm bi}-V_{\rm eq}\left(x_B-R^*\right) \label{eq:v1}\\
L &\rightarrow& L-x_B\label{eq:L1}
\end{eqnarray}

Fig. \ref{fig:ebic2d} shows the comparison between the maximum EBIC efficiency predicted by Eq. \ref{eq:2debic} and the results of the numerical simulation.  Again, there is qualitative agreement, indicating that the model identifies the key physics involved, and can accurately predict the order of magnitude required for high injection effects to occur.  The critical current density for high injection effects to occur in 2-d is given by:
\begin{eqnarray}
G_{\rm crit}^{{\rm 2d}}= \frac{\pi}{2}\mu N_A V_{\rm bi}. \label{eq:2dcrit}
\end{eqnarray}
We note the differences in the form of $G_{\rm crit}$ in 2-d as compared to the 1-d case (Eq. \ref{eq:1dcrit}).  We discuss the different forms of the critical generation rate density in various dimensions at the end of the next section.

\subsubsection{3-d analysis} \label{sec:3dmain}
Repeating the process in 3-d, we imagine a sphere of accumulated charge centered at $x_B$, initially deep within the depletion region.  Charges diffuse within the resulting screened region of radius $R^*$, while the potential drop occurs outside the screened region and ``pushes" majority carriers to the contact.  We again defer the mathematical derivation to the Appendix, and here quote the equation representing the $G=R_{\rm tot}+J$ identity:
\begin{eqnarray}
\frac{G R^*/L_D'}{\sinh\left(R^*/L_D'\right)}-\frac{2\pi\mu V_{\rm bi}N_A L}{1-R^*/L}=0 \label{eq:ebic3d}
\end{eqnarray}
The same replacements for $V_{\rm bi}$ and $L$ are used to determine the EBIC efficiency for a general beam position $x_B$ (Eqs. \ref{eq:v1} and \ref{eq:L1}).  The critical current density for high injection in 3-d is given as:
\begin{eqnarray}
G_{\rm crit}^{{\rm 3d}} = 2\pi \mu N_A V_{\rm bi} L \label{eq:3dcrit}
\end{eqnarray}

The effect of dimensionality is clearly indicated by the different dependence on the sample thickness $L$ in the expressions for the critical generation rate in 1, 2, and 3 dimensions (Eqs. \ref{eq:1dcrit}, \ref{eq:2dcrit}, and \ref{eq:3dcrit}, respectively).  In 1-d, $G_{\rm crit}$ is inversely proportional to $L$.  This is easily understood in terms of the maximum electric field that can be induced by the built-in potential $V_{\rm bi}$: as $L$ increases, the field $V_{\rm bi}/L$ decreases and the extraction rate of majority carriers is lowered.  This in turn lowers the critical generation rate.  In 2 dimensions, the total current involves an integration over the length of the contact.  The field originates from a source at the exposed surface, located a distance $L$ from the contact.  This driving field extends over a length $L$ of the contact.  This factor of $L$ from the integration cancels out the factor of $1/L$ from the electric field magnitude, leading to a critical current density which is independent of $L$.  In 3 dimensions, there is an additional spatial integration over the contact, adding another factor of $L$, so that the final critical generation density scales as $L$.

\section{experimental comparison} \label{sec:expt}

\begin{figure}[h!]
\begin{center}
\vskip 0.2 cm
\includegraphics[width=3.5in]{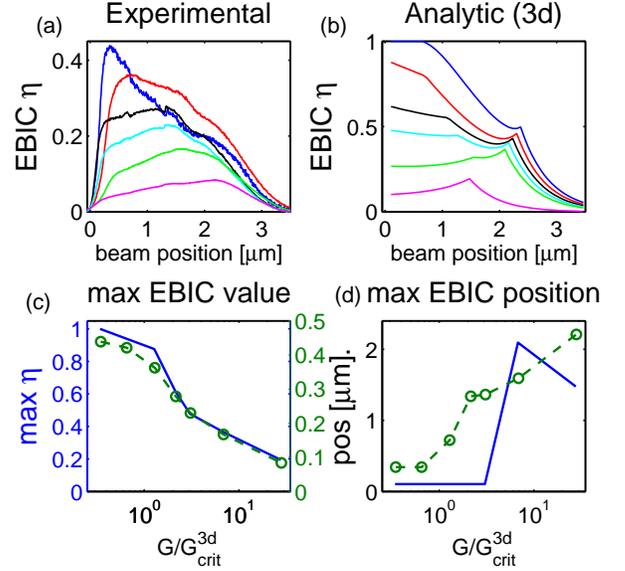}
\vskip 0.2 cm \caption{(a) shows experimental EBIC profiles for CdTe with electron beam energy of $5~{\rm keV}$, and electron beam currents of (26, 97, 162, 231, 516, and 2110) pA (in blue, red, black, cyan, green, and purple, respectively).  The uncertainty is not plotted, but is estimated as 10~\% of the EBIC efficiency.  (b) shows the analytical model lineshape for the same set of beam current values, with $\mu N_A=1.2\times 10^{14}~{\rm \left(cm \cdot V\cdot s\right)^{-1}}$, $V_{\rm bi}=1.5~{\rm eV}$, and $L_D'=500~{\rm nm}$.  (c) dashed green curve shows the experimental maximum EBIC efficiency as a function of total generation rate (scaled by critical generation rate, taken experimentally from letting $I_{\rm crit}=75~{\rm pA}$).  Solid blue line is the same result for the analytical model.  Note the experimental (analytic) y-axis is on the left (right).  (d) dashed green curve shows the experimental position of the EBIC maximum, while the solid blue line is the analytical model.}\label{fig:ebic3d2}
\end{center}
\end{figure}

We make two comparisons between the 3-dimensional model and experimental data.  We present EBIC data from two rather different samples.  We first consider a device with power conversion efficiency of 10 \% and a nominal CdTe thickness of $3~{\rm \mu m}$.  We prepare cross sectional samples by cleaving the device, in order to minimize the effects of surface damage by additional processes such as focused ion beam milling.  We present EBIC lineshape taken from a large grain (about $1~{\rm \mu m}$) in order to minimize the effect of grain boundaries.  Fig. \ref{fig:ebic3d2} shows the experimental EBIC signal for six values of beam current, at a fixed beam energy of $5~{\rm keV}$ (resulting in an excitation with length scale of about $100~{\rm nm}$).  We estimate 10~\% relative uncertainty in the measured EBIC efficiency $\eta$ (all uncertainties are reported as one standard deviation).  The dominant sources of uncertainty are from the beam current, and from the inhomogeneous material composition, which introduces uncertainty into the the backscattering coefficient $b$ in Eq. \ref{eq:G} (performed for pure CdTe).  The length scale for this material inhomogeneity ({\it e.g.} alloying) is much smaller than the electron beam spot size, so that the error is uniform across linescans.  The uncertainty is therefore in the magnitude of the EBIC signal, not the shape.  We estimate an uncertainty in the maximum position of $50~{\rm nm}$ based on the discretization of the electron beam position in the linescan.

Qualitatively, for increasing beam current, the maximum EBIC signal is decreased, and the profile is significantly broadened.  This is in agreement with the model, as shown in Fig. \ref{fig:ebic3d2}(b) (see caption for model parameters).  We find a discrepancy between absolute values of EBIC: the experimental EBIC is always much lower than 1.  We attribute the overestimation of the EBIC magnitude in the model to other effects which reduce the lifetime, such as surface effects, or high injection effects distinct from those considered in this work (described in detail in Ref. \cite{haney}).  Fig. \ref{fig:ebic3d2}(c) shows the experimental and model prediction for the maximum EBIC value as a function of the total generation rate.  We find good agreement in the overall trend, despite a relatively constant reduction of 0.6 of the experimental EBIC efficiency relative to the model result (note the different y-axes).  Fig. \ref{fig:ebic3d2}(d) shows the position of the maximum EBIC value.  The agreement is qualitatively reasonable, with discrepancies likely arising from the approximations involved in formulating the analytical model (which lead to predicted EBIC curves with two kinks, as discussed earlier).

\begin{figure}[h!]
\begin{center}
\vskip 0.2 cm
\includegraphics[width=3.5in]{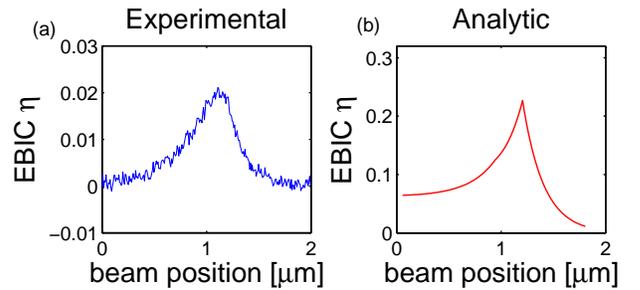}
\vskip 0.2 cm \caption{(a) Experimental EBIC curve of device exhibiting 3 \% power conversion efficiency, with electron beam energy of 10 keV and current 300 pA.  (b) Theoretical EBIC from Eq. \ref{eq:ebic3d}, using parameters for resistivity as determined from the device J-V curve $\mu N_A=3\times 10^{13}~{\rm \left(cm\cdot V \cdot s\right)^{-1}}$, $V_{\rm bi}=1.4~{\rm eV}$, $L_D'=200~{\rm nm}$.}\label{fig:ebic3d1}
\end{center}
\end{figure}

We next consider a different sample with very different device properties.  It demonstrates a low power conversion efficiency of 3 \% and a thinner CdTe layer with thickness $2.0~{\rm \mu m}$.  The EBIC lineshape at a beam energy of $10~{\rm keV}$ and beam current of $300~{\rm pA}$ is shown in Fig \ref{fig:ebic3d1} (a).  The experimental lineshape is the average over many grains; in general we find moderate to low variation in the lineshape across different grains for this sample.

To make comparisons with the model, we first note that this device exhibits a substantial series resistance-area product $R_SA$ (which is partially responsible for its low efficiency).  Fitting the J-V curve to an equivalent circuit model, we find $R_SA\approx 2\times10^{-3}~{\rm {\Omega\cdot m^2}}$.  The contribution from the absorber to the series resistance is given by $\rho/\left(L-L_W\right)$, with resistivity $\rho=\left(q\mu N_A\right)^{-1}$, and where $\left(L-L_W\right)$ is the thickness of the neutral region.  We can thus make an estimate of $\mu N_A=\left(L-L_W\right)/\left(qR_SA\right)=3\times 10^{13}~{\rm \left(cm\cdot V\cdot s\right)^{-1}}$.  The corresponding factor in Eq. \ref{eq:ebic3d} is fixed to this value, leaving the effective diffusion length $L_D'$ and $V_{\rm bi}$ as free parameters.  The model EBIC profile is plotted in Fig. \ref{fig:ebic3d1}(b) for $L_D'=200~{\rm nm}$.  We find a lineshape which is very similar to the experiment, although larger in magnitude, as discussed for the previous sample.  Despite the discrepancy in absolute value, the similar lineshape feature of a maximum position in the middle of the absorber layer is a strong indication that the model presented here is relevant for samples with a high bulk resistivity.  Indeed, Eq. \ref{eq:3dcrit} can be reformulated in terms of the serial resistance-area product:
\begin{eqnarray}
G_{\rm crit}^{{\rm 3d}} = \frac{2\pi  V_{\rm bi} L^2}{R_S A} \label{eq:3dcrit2}
\end{eqnarray}
Given an estimate of $R_SA$, Eq. \ref{eq:3dcrit2} provides a simple expression for the onset of high injection screening effects for low energy EBIC experiments.

\section{Conclusion}
In this work, a model is developed to describe the photovoltaic response to a point source excitation in the high injection regime.  This situation is most relevant to EBIC experiments on materials with high resistivity.  The important outcomes of the model include a simple expression for the threshold excitation rate, above which high injection effects occur, and a straightforward procedure for computing the EBIC lineshape in the high injection regime.  The signature of high injection and screening effects are a reduced maximum EBIC collection efficiency, and an EBIC lineshape which is broadened near the junction, and whose maximum value may be positioned away from the depletion region.  These features are observed for CdTe solar cells probed with electron beam currents typical for EBIC experiments.  The model enables an interpretation of EBIC signals for materials with high resistivity, so that the high spatial resolution advantage of EBIC may be leveraged to extract quantitative information about these materials at the nanoscale.

\section*{Acknowledgment}
H. Y. acknowledges support under the Cooperative Research Agreement between the University of Maryland and the National Institute of Standards and Technology Center for Nanoscale Science and Technology, Award 70NANB10H193, through the University of Maryland.  P. K. and R. W. C. were supported by the DOE/NSF F-PACE Program (Contract DE-EE0005405).

\begin{appendix}

\section{1d} \label{app:1d}
We describe in more detail the mathematics of the 1-d analytical model.  We assume that the charges accumulate at the excitation position $x_B$ and screen the built-in field.  For a delta-function excitation of magnitude $G$, the boundary conditions within the screened region are:
\begin{eqnarray}
p_1\left(x_B\right) - p_2\left(x_B\right)  &=& 0\\
j_1\left(x_B\right) - j_2\left(x_B\right) &=& G \\
p_2\left(L^*\right) &=& 0, \label{eq:bc1}\\
j_1\left(x_m\right) &=& 0. \label{eq:bc2}
\end{eqnarray}
where $j=-D\partial p(x)/\partial x$, and the $1~(2)$ subscript indicates the solution to the left (right) of the excitation position $x_B$.  The vanishing charge density at the right edge of the screened region, specified in Eq. \ref{eq:bc1}, is a consequence of charges being swept out by the field there.  The vanishing current at the left edge of the screened region (Eq. \ref{eq:bc2}) follows from the assumption that the contact at $x=0$ collects electrons and blocks holes.  Note that the position of the left edge of the screened region is denoted by $x_m$, and is given by
\begin{eqnarray}
x_m=\max\left(0,x_B-L^*\right).
\end{eqnarray}
Physically, for excitations near the $p$-$n$ metallurgical junction, the boundary condition at $x_m$ applies at the physical edge of the device $x=0$.  For excitations more than a screening length $L^*$ away from the junction, the boundary condition at $x_m$ applies at the ``back" edge of the screened region $x_B-L^*$.

The resulting carrier density is given by:
\begin{eqnarray}
p(x)= A\times\left\{
\begin{array}{rl}
\cosh\left(\frac{x-x_B+x_m}{L_D'}\right)\sinh\left(\frac{L^*}{L_D'}\right) & {\rm if}  x < x_B,\\
\sinh\left(\frac{L^*-x+x_B}{L_D'}\right)\cosh\left(\frac{x_m}{L_D'}\right) & {\rm if}  x \geq x_B,
\end{array} \right.
\end{eqnarray}
where $A=\left(G L_D'/D\right)\times {\rm sech}\left[\left(x_m+L^*\right)/L_D'\right]$.  As described in the main text, the total recombination is given by the spatial integral of $p(x)/\left(2\tau\right)$, with the result given in Eq. \ref{eq:R1d} in the main text.

\section{2d} \label{app:2d}
The model to describe the 2 dimensional system is depicted in Fig. \ref{fig:2d}(a).  We suppose that the built-in field is screened in the vicinity of the excitation, taking the screening length to be $R^*$.  Within this region, charges diffuse and recombine.  To keep the mathematics analytically tractable, we assume complete radial symmetry for the screened region in which charges diffuse, ignoring the surface entirely.  The radial diffusion equation is:
\begin{eqnarray}
\frac{1}{r}\partial_r p + \partial_r^2 p = \frac{p}{2D\tau}
\end{eqnarray}
The boundary conditions are a point source excitation at $r=0$ together with vanishing minority carrier density at $r=R^*$:
\begin{eqnarray}
-\lim_{r\rightarrow 0}2\pi r D \partial_r p &=& G \\
p(R^*/L_D') &=& 0
\end{eqnarray}
The solution is a linear combination of 0th-order Bessel functions $J_0$ and $Y_0$ with imaginary arguments:
\begin{eqnarray}
p(r) &=& A J_0\left(ir/L_D'\right) + B Y_0\left(-ir/L_D'\right)
\end{eqnarray}
The constants $A$ and $B$ which satisfy the boundary conditions are given as:
\begin{eqnarray}
A&=&\frac{G }{4 D } \frac{Y_0\left(-iR^*/L_D'\right)}{J_0\left(iR^*/L_D'\right)}\\
B&=&\frac{-G}{4 D}
\end{eqnarray}
The total recombination is the integral of the carrier density over the screened region \cite{footnote3}:
\begin{eqnarray}
R_{\rm tot} &=&  \int_0^{R^*} 2 \pi r ~ \frac{p\left(r\right)}{2\tau}~ dr \\
&=& \frac{-2\pi G L_D'^2}{4 \tau D} \left(\frac{iR^*}{L_D'}\frac{Y_0\left(-iR^*/L_D'\right)}{J_0\left(iR^*/L_D'\right)} J_1\left(iR^*/L_D'\right)  \right.\nonumber \\ &&\left. ~~~+\frac{iR^*}{L_D'}Y_1\left(-iR^*/L_D'\right) + \frac{2i}{\pi}\right)\label{eq:R2d}
\end{eqnarray}

Next we estimate the electric field induced by the distortion of the potential from the charge screening.  The total current is an integral over the length of the contact:
\begin{eqnarray}
J &=& \int dz J_x(z) =\mu N_A \int dz E_x(z)
\end{eqnarray}
To evaluate $E_x(z)$, we make the ansatz that the potential at the edge of the screened region is $V_0$.  This is effectively the ``source" of the nonequilibrium electric field which drives majority carriers out through the contact at $x=L$.  We approximate the magnitude of the field at position $z$ to be $V_{\rm bi}/\left(\sqrt{L^2+z^2}-R^*\right)$ - this amounts to assuming that the potential drops linearly from the edge of the screening region to the contact.  The integral over $z$ is conveniently performed using the variable $\theta$, shown in Fig. \ref{fig:2d}(a).
\begin{eqnarray}
J &=&\mu N_A V_{\rm bi}\int_{-\pi/2}^0 d\theta \frac{1}{1-R^*\cos\left(\theta\right)/L}\\
&=&\mu N_A V_{\rm bi} \frac{2}{\sqrt{R^*/L-1}}\tanh^{-1}\left(\frac{1+R^*/L}{\sqrt{R^*/L-1}}\right) \label{eq:J2d}
\end{eqnarray}
Equating the sum of Eqs. \ref{eq:R2d} and \ref{eq:J2d} to the total generation rate leads to Eq. \ref{eq:2debic} of the main text.

\section{3d} \label{sec:3d}
We present the same analysis for the 3-dimensional case.  We again make the simplifying assumption that the screened region is spherically symmetric, and take the minority carrier density to vanish at the edge of the screened region (which has radius $R^*$).  As before, the carrier generation results in a discontinuity in the radial current at the injection point $r=0$:
\begin{eqnarray}
\lim_{r\rightarrow 0} -4\pi D \partial_r p\left(r\right) &=& G\\
p(R^*) &=&0.
\end{eqnarray}
The resulting density is:
\begin{eqnarray}
p\left(r\right) = \frac{G}{D 4 \pi r}\frac{\sinh\left(\frac{R-r}{L_D'}\right)}{\sinh\left(\frac{R}{L_D'}\right)}
\end{eqnarray}
This leads to the total recombination:
\begin{eqnarray}
R_{\rm tot} &=& \frac{4\pi}{2\tau} \int_0^{R^*} r^2 dr~ p\left(r\right) \\
&=&G   \left(   \frac{\sinh\left(R^*/L_D'\right) - R^*/L_D' }{\sinh\left(R^*/L_D'\right)}    \right) \label{eq:R3d}
\end{eqnarray}

\begin{figure}[h!]
\begin{center}
\vskip 0.2 cm
\includegraphics[width=3.in]{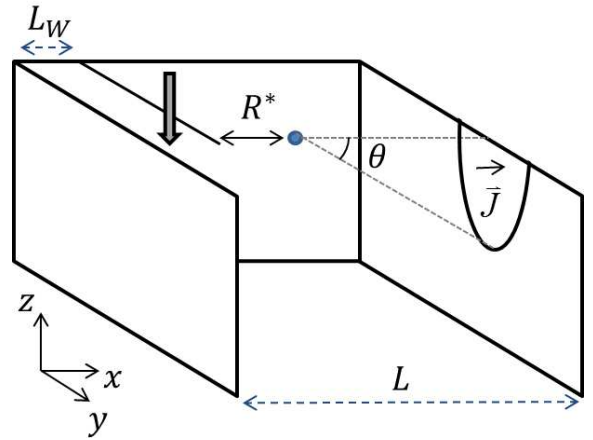}
\vskip 0.2 cm \caption{Cartoon of the model to describe the high injection regime in 3 dimensions.}\label{fig:3d}
\end{center}
\end{figure}

For the total collected current, we integrate over $z$ and $y$ (see Fig. \ref{fig:3d}:
\begin{eqnarray}
J &=& \mu N_A \int  E_x(y,z)dz dy  \\
J &=& 2\pi \mu N_A V_{\rm bi} L \int_{-\pi/2}^0 d\theta \frac{\sin\left(\theta\right)}{\left(1-R^*\cos\left(\theta\right)/L\right)^2}\\
 &=& \frac{2\pi \mu N_A V_{\rm bi}L}{\left(R^*/L\right)-1} \label{eq:J3d}
\end{eqnarray}

Equating the sum of Eqs. \ref{eq:R3d} and \ref{eq:J3d} to the total generation rate leads to Eq. \ref{eq:ebic3d} of the main text.

\end{appendix}


\begin{thebibliography}{1}
\bibitem{footnote1}
To convert the beam current to a generation rate in a 1-d model, we divide the 3-d generation rate (units of ${\rm s^{-1}}$) by $L_D'^2$, so that the generation rate units are appropriate for a 1-d model.
\bibitem{footnote2}
For Shockley-Read-Hall trap-mediated recombination, the hole lifetime increases from $\tau$ when $p \ll n$ to $2\tau$ when $p=n$.  When $p \ll n$, localized states are almost exclusively occupied by electrons, so that the hole recombination lifetime is inversely proportional to the trap density - in a trap, a hole will always find an electron to recombine with.  However, for $p=n$, there is an equal occupation of traps by electrons and holes, so that a hole will find an electron to recombine with only half the time.  This increases the hole lifetime by a factor of 2.
\bibitem{footnote3}
The presence of the surface suggest that the angular integral should be performed between $\theta=0$ and $\theta=\pi$.  However, we find the result conforms with the numerical model if the integral is performed over the entire circle of the screened region.  This is because our assumption of a circular boundary condition (which ignores the surface) in Eq. leads to lower charge density in the lower half circle (and therefore lower recombination), relative to the value that would be obtained with a proper surface boundary condition.  Integrating over the entire circle increases the total recombination, partially compensating for this initial underestimation.
\end{thebibliography}
\end{document}